\documentclass[sigconf]{acmart}
\usepackage{lipsum}

\usepackage{graphicx}

\usepackage[disable]{todonotes}
\usepackage{booktabs}
\PassOptionsToPackage{hyphens}{url}\usepackage{hyperref}

\usepackage{dsfont}
\usepackage{amsmath}
\usepackage{bbm}
\usepackage{xfrac}

\newcommand{\E}[1]{\mathbb{E}\left[#1\right]}

\usepackage{caption}
\usepackage{subcaption}

\usepackage{xcolor}

\AtBeginDocument{%
  \providecommand\BibTeX{{%
    \normalfont B\kern-0.5em{\scshape i\kern-0.25em b}\kern-0.8em\TeX}}}

\copyrightyear{2021}
\acmYear{2021}
\setcopyright{acmlicensed}\acmConference[DocEng '21]{ACM Symposium on Document Engineering 2021}{August 24--27, 2021}{Limerick, Ireland}
\acmBooktitle{ACM Symposium on Document Engineering 2021 (DocEng '21), August 24--27, 2021, Limerick, Ireland}
\acmPrice{15.00}
\acmDOI{10.1145/3469096.3469873}
\acmISBN{978-1-4503-8596-1/21/08}

\begin{document}

\title{Heuristic Stopping Rules For Technology-Assisted Review}

\author{Eugene Yang}
\affiliation{%
  \institution{IR Lab, Georgetown University}
  \city{Washington}
  \state{DC}
  \country{USA}
}
\email{eugene@ir.cs.georgetown.edu}

\author{David D. Lewis}
\affiliation{%
  \institution{Reveal Data}
   \city{Chicago}
   \state{IL}
  \country{USA}
}
\email{sigir2021paper@davelewis.com}

\author{Ophir Frieder}
\affiliation{%
  \institution{IR Lab, Georgetown University}
  \city{Washington}
  \state{DC}
  \country{USA}
}
\email{ophir@ir.cs.georgetown.edu}

\renewcommand{\shortauthors}{Yang, Lewis, and Frieder}

\begin{abstract}
Technology-assisted review (TAR) refers to human-in-the-loop active learning workflows for finding relevant documents in large collections.  These workflows often must meet a target for the proportion of relevant documents found (i.e. recall) while also holding down costs.  A variety of heuristic stopping rules have been suggested for striking this tradeoff in particular settings, but none have been tested against a range of recall targets and tasks.  We propose two new heuristic stopping rules, \textit{Quant} and \textit{QuantCI} based on model-based estimation techniques from survey research. We compare them against a range of proposed heuristics and find they are accurate at hitting a range of recall targets while substantially reducing review costs. 

\end{abstract}




\maketitle

\section{Introduction}

\textit{Technology-assisted review (TAR)} refers to human-in-the-loop iterative active learning workflows for large scale document review. A major application area is  eDiscovery: review of documents in the law (civil litigation, regulatory review, and investigations)~\cite{baron2016perspectives,totalrecall2015,totalrecall2016}. Other applications include  open government document requests~\cite{baron2020providing} and systematic review in medicine \cite{wallace2010semi,clef2019ehealth-tar, clef2018ehealth-tar, clef2017ehealth-tar}. 

On each iteration a TAR workflow uses a predictive model to select a batch of documents to review (typically a few hundred) using an active learning method such as relevance feedback~\cite{rocchio1971relevance,lewis1994sequential,cormack2014evaluation}. Those documents are reviewed by a human expert and added to the collection of reviewed documents. The review documents are used to train another predictive model, and the cycle repeats. 

In a \textit{one-phase TAR workflow} this process is iterated until some stopping condition is met. Relevance feedback (training on top-ranked documents) is often used as the active learning method.   In a \textit{two-phase TAR workflow}, the process is stopped before completion and the final trained classifier is used to identify a large set of documents to go to another review team (often at a lower per-document cost) to finish the review~\cite{cost-structure-paper}. 

TAR workflows typically must meet a target for the proportion of relevant documents found (i.e. recall) while also holding down costs.  The choice of stopping point is therefore critical. Two-phase workflows often condition stopping on a sample-based estimate of recall, with the cost of random sampling offset by the cost savings from using a cheaper second phase review team.  One-phase reviews, on the other hand, are often used in situations, e.g. a single investigator searching for key information, where the cost of labeling a sample purely to support a stopping rule is untenable.  

A variety of heuristic stopping rules have therefore been proposed for stopping one-phase reviews~\cite{li2020stop, callaghan2020statistical, cormack2016engineering, cormack2016scalability, totalrecall2015, totalrecall2016}.  These have mostly been developed in the context of a single TAR application or recall goal and their general applicability is unknown.  

In this paper we propose the theory of model-based estimation in survey research  \cite{sarndal1978design} provides a foundation for more flexible and rigorous stopping rules in technology-assisted review. Our contributions are (1) two new heuristic stopping rules, \textit{Quant} and \textit{QuantCI}, that adjust for task characteristics and let stopping be conditioned on an arbitrary user-chosen recall goal, and (2) a rigorous evaluation of our proposed rules against heuristic stopping rules from both the academic literature and current industry practice.  We examine how the recall goal, prevalence of relevant documents, and the difficult of the classification problem combine to affect stopping rule behavior. We  find our proposed rules are accurate at hitting a range of recall targets while substantially reducing review costs.

\section{Background}
\label{sec:background}

We can classify stopping rules for TAR workflows into three groups: sample-based, interventional, and heuristic.  Our focus in this paper is on one-phase reviews, so we emphasize stopping rules for those.    

\textit{Sample-based stopping rules} are based on drawing a random sample from a collection of interest, having it reviewed, and using that labeled sample to estimate the recall of review process at each potential stopping point. Sample-based stopping rules have been proposed for both one-phase~\cite{qbcb-paper, cormack2016engineering, saha2015batch-mode} and two-phase~\cite{bagdouri2013towards, webber2013sequential} reviews.  When estimated recall is high enough, the review is stopped or, in the case of a two-phase review, a cutoff is set on the second phase. 

Sample-based stopping rules are the method of choice when a statistical guarantee on effectiveness is needed, as in some legal contexts.  Unfortunately, the size of a sample necessary to estimate recall with a given confidence is inversely proportional to the prevalence of relevance documents~\cite{lewis2016defining}. For example, estimating recall with a margin of error of 0.05 and a confidence level of 95\% requires a simple random sample containing roughly 385 relevant documents.  If the prevalence of relevant documents was 1\%, one could need on average a sample size of 38,500 documents. This makes sample-based rules prohibitively expensive in many settings. 

\textit{Interventional stopping rules} instead take the approach of modifying the TAR process itself in support of estimating recall.  One approach is to modify the active learning algorithm to incorporate probabilistic sampling~\cite{cormack2016scalability,li2020stop}. This lets the selected data do triple duty: training the selection model, estimating recall, and accomplishing review. Another approach is to terminate conventional active learning at some point and switch to purely random review~\cite{callaghan2020statistical}, allowing recall to be estimated. While powerful, these methods are not a free lunch: they reduce the effectiveness of active learning in exchange for supporting estimation. They also cannot be used, for instance, by review managers who are limited to the active learning methods provided by a particular  piece of commercial review software. 

\textit{Heuristic stopping rules} are the most generally applicable, and are the focus of this paper.  They make a decision of when to stop a TAR workflow using only the data naturally labeled during the workflow.  We can distinguish between \textit{target-aware} heuristic rules which modify their behavior based on a desired recall target, and \textit{target-agnostic} heuristic rules that attempt to optimize some general condition on the quality of the review. 
We survey a wide range of particular heuristic rules in the next section.

\section{Heuristic Stopping Rules}
\label{sec:existing-rules}

In this section we review a range of heuristic stopping rules from both the research literature and operational practice. While our treatment is not exhaustive, we cover the major styles of rules that have been proposed.

\subsection{Fixed Iterations Method}

Both one-phase and two-phase TAR processes are sometimes stopped simply after a predetermined number of training iterations as a ``\textit{simple, operational stopping criteria}'' described by \citet{wallace2010semi}. Other works also referred this rule as a \textit{pragmatic stopping rule}~\cite{callaghan2020statistical}. This has the advantage of making review cost known in advance (assuming uniform document review costs), but provides no ability to adapt to the category prevalence, difficulty, or recall target. 

\subsection{The ``2399'' Method}

Some early work in document review for eDiscovery used stopping rules based on a sample-based estimate of elusion (the proportion of relevant documents in the unreviewed population) ~\cite{roitblat2007search}. A sample of 2399 documents is sufficient to estimate elusion with 95\% confidence and a 2\% margin of error. This led to an odd  fixation on the number 2399 in eDiscovery. 

One result was the the ``2399'' heuristic method~\cite{cormack2015waterloo, cormack2016scalability}.  
It stops a one-phase TAR review after $2399 + xR$ documents have been reviewed,  where $R$ is the number of positive documents the review  has found, and $x$ is a tuned hyperparameter.  Our experiments use $x=1.2$ as proposed by the rule's inventors~\cite{cormack2015waterloo, cormack2016scalability}. 

\subsection{Batch Positives Method}

A more adaptive approach to one-phase TAR reviews is to stop iterations when one or more recent training batches contain only a few positive documents, i.e., low precision.  This can be given an economic rationale. If the fraction of relevant documents in a batch is $x$, then the number of documents that must be examined to find a single relevant document is roughly $1/x$.  This is a measure of the marginal cost to find the next relevant document.  Such marginal utility arguments can be linked to the legal concept of  proportionality in the discovery process~\cite{hirt2011applying}.  Further, under the assumption that batch precision declines in a roughly monotonic fashion after some point in training, once the marginal cost exceeds a threshold, it never declines, so stopping is appropriate.

\subsection{Probability Cutoff Method}

Closely related to the batch precision rule are rules that stop review when all unreviewed documents have a score below some cutoff.  In particular, assume that the scores are in fact well-calibrated predicted probabilities. Then the reciprocal $1/p$ of a probability score $p$ is the number of documents that must be examined at that point to find a single relevant document. As with batch precision, this is a measure of the marginal cost of finding the next relevant document. However, also as with batch precision, there is no particular connection with recall goals.

\subsection{Knee Method}

A %
refinement of the Batch Positive method is the \textit{Knee Method} \cite{cormack2016engineering}. This is based on the gain curve \cite{totalrecall2015} for a one-phase TAR process.  A gain curve plots how the number of positive documents found increases as more documents are reviewed, typically on a per-batch basis.

At each potential stopping point $s$, the knee method computes the ratio of slopes in a two-segment approximation to the gain curve: 
\begin{eqnarray}
   \rho(s) 
      &=& \frac{Slope( (0, 0), 
                       (i, Rel(i)) ) 
               }{
                Slope( (i, Rel(i)), 
                       (s, Rel(s) + 1)
               } \nonumber \\ 
      &=& \frac{Rel(i)}{i} \frac{s-i}{Rel(s) - Rel(i) + 1}
\end{eqnarray}

Here $Rel(k)$ is the number of relevant documents found at rank k. The value of $i$ is chosen such that $(i,Rel(i))$ is a ``knee'', i.e., has maximum perpendicular distance from the line segment between $(0,0)$ and  $(s,Rel(s))$.

The Knee Method stops at the first $s$ such that

\begin{itemize}
      \item $\rho(s) \ge 156 - min(Rel(s), 150)$, and  
      \item $s \ge 1000$
\end{itemize}

The Knee Method is targeted at a recall goal of 0.7, and \citet{cormack2016engineering} do not discuss how it might be adapted to other recall targets. 

\subsubsection{Adapting the Knee Method to Fixed Size Batches}

\citet{cormack2015autonomy} specify that values of $s$ for the Knee Method should be based on the batch schedule used by their BMI (Baseline Model Implementation) system\footnote{\url{https://plg.uwaterloo.ca/~gvcormac/trecvm/}}. That schedule uses a single relevant seed document (batch size $B = 1$) on iteration 1\footnote{For supervised learning, 100 random documents artificially labeled as negative examples are used on round 1 only, but these are not considered to be part of the batch size.}. Then the batch size $B_k$ for round $k$ is $B_{k-1} + \lceil B_{k-1}/10 \rceil$, i.e., batches grow exponentially in size.

\begin{table}[t]
\centering
\caption{Transformed rounds for knee tests from exponential scheduling in BMI to fixed batch size of 200.  }
\label{tab:BMI-schedule}

\begin{tabular}{cccc|ccc}
 \toprule
 \multicolumn{4}{c|}{BMI Schedule} &
 \multicolumn{3}{c}{Batch Size 200 Schedule} \\
 \midrule
        & Batch & Training  & Knee  &       & Training &  Knee  \\
  Round &  Size &  Set Size & Test? & Round & Set Size &    Test?   \\
  \midrule
        &       &              &     &    5   &    801   &  No   \\
    33  &  104  &      991     &  No  &   6    & 1001    &  No     \\
    34  &  115  &     1106     &  Yes  &  7    & 1201    &  Yes      \\
    35  &  127  &     1233     &  Yes  &       &          &     \\
    36  &  140  &     1373     &  Yes  &  8    &  1401    &  Yes      \\
    37  &  154  &     1527     &  Yes &   9   & 1601     &  Yes \\
    38  & 170   &     1697     & Yes  &  10   & 1801     &  Yes \\
    39  & 187  &     1884     & Yes  &  11    & 2001     & Yes  \\
    40  &  206 &     2090     &  Yes &  12    &  2201    & Yes \\
    41  & 227 &      2317     &  Yes & 13     &  2401      & Yes  \\
    42  & 250 &      2567     &  Yes & 14     &  2601       &  Yes \\
        &     &               &      & 15     &  2801    & No \\
    43  & 275 &      2842     & Yes & 16      &  3001      & Yes \\
  \bottomrule
\end{tabular}    
\end{table}

We can easily adapt the Knee Method, however, to fixed size batches. Since the method is intended to be conservative, our goal is to never stop sooner than the Knee Method would, but also to bound the extra cost we incur by using fixed size batches. We assume, as in the BMI schedule, that the first round is a seed set consisting of a single document, but then follow that with fixed size batches.  

We assume batches of size 200 as an example. Table~\ref{tab:BMI-schedule} shows the BMI batch schedule in the vicinity of the earliest potential stopping point: $s=1000$. For batch size 200, the first fixed training set size over 1000 is 1001. Since that is smaller than the first Knee Method batch, we conservatively wait until training set size 1201 for the first check on the modified schedule.  We then check at training set sizes 1401, 1601, ... 2601, each of which is subsequent to at least one check that the BMI-based knee method would do. We do not check at 2801, because it would not be conservative to give an additional chance to stop then.  We do check at training set size 3001 (greater than 2842), and subsequently at the first training set size larger than each training set size for the BMI-based schedule.

\subsection{Budget Method}

\citet{cormack2016engineering} also proposed the \textit{Target Method}, which draws a random sample of documents until 10 positive documents are found, hiding those documents from the TAR process.  It then stops a one-phase TAR review after the 10 positive documents are rediscovered~\cite{cormack2016engineering, qbcb-paper}. As a sample-based method, it is out of the scope of this paper, but \citet{cormack2016engineering} have proposed a heuristic alternative to it: the Budget Method.  

The Budget Method is based on three observations: %

\begin{itemize}
    \item For any amount of effort, a TAR process should find more positive documents than would random sampling.  
    \item The target method on average will draw a target set of roughly $10C/R$ documents, where $C$ is the collection size and $R$ is the number of relevant documents in the collection.
    \item At any moment in time, the number of relevant documents $Rel(s)$ is a lower bound on $R$, and thus $C/Rel(s)$ is an upper bound on the expected size of the size of the sample the target method would draw.  
\end{itemize}

The Budget Method stops review at the first end-of-batch training set size $s$ if 
\begin{itemize}
    \item $s \ge 0.75C$, or
    \item both $\rho(s) \ge 6$ and $s \ge 10C/Rel(s)$ are true
\end{itemize}

The first test is based on the assumption that reviewing a 75\% random sample of the collection would likely achieve a recall of at least 0.7, and reviewing 75\% found by a TAR process should do even better.  The second way of stopping is based on two criteria: a simplified version of the knee test that is assumed to be safe when $R$ is large, and a test that keeps the knee test from taking effect until $Rel(s)$ is somewhat large.  Again, it is not clear how to adapt this method to recall goals differing 0.7.

\subsection{CMH Heuristic Method}

Recently, \citet{callaghan2020statistical} proposed an interventional method that stops the TAR process based on a heuristic method, and then incrementally samples random documents to both fulfill the recall target and statistically verify the fulfillment. While the sampling phase is interventional (and thus out of scope for this paper), the  heuristic approach (\textit{CMH heuristic method}) is of  interest. 

At each round of active learning with end-of-batch training set size $s$, the CMH heuristic method splits the successive rounds into two parts by a pivot round $j$, hypothetically assumes the second part was sampled randomly, and calculates the probability of the current recall being larger than or equal to the target recall $p_j$. Specifically, let $s_j$ being the end-of-batch training set size of any successive round $j$ and $\tau_{tar}$ being the recall target,
\begin{align}
    & p_j = P\Big(X \le Rel(s) - Rel(s_j) \Big) \nonumber\\
    & \text{where } X \sim \text{Hypergeometric}(C-Rel(s_j), K_{tar}, s-j) \nonumber \\
    & \phantom{\text{where }} K_{tar} = \Bigg\lfloor \frac{Rel(s)}{\tau_{tar}} - Rel(s_j) + 1 \Bigg\rfloor \nonumber
\end{align}

The TAR process is stopped if $min(p_j) < 1 - \alpha$ for some confidence level $\alpha$. In our experiment, we follow the original proposal of the method and set $\alpha$ to 95\%. 
\section{Stopping Based on a Recall Estimate}

The heuristic stopping rules above except the CMH heuristic method either ignore the review's recall target, or are designed for a particular target (e.g. 0.7 for the knee rule). A practical heuristic stopping rule should allow an arbitrary recall target to be chosen and respond accordingly. 

Stopping based on an estimate of recall is a natural approach. Labeling a random sample purely to estimate recall is often viewed as too expensive (Section~\ref{sec:background}). However, if recall could be estimated, even roughly, from the training data already chosen by relevance feedback this would provide a broadly applicable rule. We present such an approach below based on the statistical technique of model-based estimation.

\subsection{Model-Based Estimation}

Using randomness properties of a sample to justify extrapolation to a population (\textit{design-based estimation}) \cite{sarndal1978design} is not the only approach to estimation. An alternative, widely used in survey research, is \textit{model-based estimation} \cite{sarndal1978design}. In this approach, (1) the sampled items are used to fit a predictive model in a supervised fashion, (2) that model is applied to unsampled items to predict a value for each, and (3) the predicted values for unsampled items are used in producing the population estimate.  In information retrieval, similar approaches have been explored for text quantification tasks \cite{lewis1995evaluating,thomas1995text,gao2016classification}.

In our context, we already have an appropriate predictive model: the logistic regression model (Section~\ref{subsec:implementation}) trained on relevance feedback samples to prioritize documents for review. It outputs, for any document $i$, a predicted probability of relevance $p_i$, i.e. an estimate of a 0/1 relevance label.  If $\mathcal{R}$ and $\mathcal{U}$ are the set of reviewed and unreviewed documents respectively, then 
\begin{eqnarray}
    \widehat{R}_r &=& \sum_{i\in \mathcal{R}} p_i \label{eq:model-based-estimate-of-relevant-in-reviewed} \\
    \widehat{U}_r &=& \sum_{i\in \mathcal{U}} p_i \label{eq:quant-est}
\end{eqnarray}
are, respectively, the model-based estimate of the number of relevant documents in the reviewed $\mathcal{R}$ and unreviewed documents $\mathcal{U}$. Two plausible point estimates of recall are then:  
\begin{eqnarray}
    \frac{R_r}{ R_r + \widehat{U}_r } 
           & = & 
           \frac{R_r}{ R_r + \sum_{i\in \mathcal{U}} p_i }
    \label{eq:quant-recall-using-true-reviewed-relevant} 
         \\
    \frac{\widehat{R}_r}{\widehat{R}_r + \widehat{U}_r} 
            & = & 
            \frac{
                  \sum_{i \in \mathcal{R}} p_i
                  }{ 
                   \sum_{i\in \mathcal{C}} p_i
                  }
    \label{eq:quant-recall-using-estimated-reviewed-relevant}
\end{eqnarray}
Equation~\ref{eq:quant-recall-using-estimated-reviewed-relevant} may seem strange, since we know $R_r$ (the number of reviewed relevant documents). However, recall is a ratio estimator, and there is an advantage to having any systematic modeling error affecting $\widehat{U}_r$ to be present in both numerator and denominator. We in practice found Equation~\ref{eq:quant-recall-using-estimated-reviewed-relevant} to provide substantially better results, so discuss rules based only on that.

This estimate is based on strong assumptions that rarely hold exactly. In this case the assumption is that the probabilities from the relevance model are well-calibrated, i.e., that if one took a large sample of documents with predicted probability $p_i$, it would contain a fraction $p_i$ of relevant documents \cite{dawid1982well}.

\subsection{Approximating Variance}

Stopping when a point estimate, even an unbiased one based on a random sample, equalled the recall goal would miss that goal a substantial fraction of the time. (50\% of the time if sampling errors fell equally on either side of the point estimate.) Stopping rules based on sample-based estimates address this by instead stopping when the lower bound of a confidence interval exceeds the recall goal \cite{qbcb-paper,cormack2016engineering,bagdouri2013towards}. As larger samples are used, the confidence interval narrows and stopping occurs sooner. 

Producing a confidence interval requires an estimate of the variance of the point estimate. A general method for approximating the variance of a complicated random variable (in our case our recall estimator) is to use a truncated Taylor series.  This is variously referred to as the \textit{linearization method}, \textit{propagation of error method}, \textit{delta method}, and \textit{Taylor series method} \cite{valliant2013practical, webber2013approximate}.

The estimate in Equation~\ref{eq:quant-recall-using-estimated-reviewed-relevant} is based on modeling the relevance of a document $i$ as the outcome of a Bernoulli random variable $D_i \sim Bernoulli(p_i)$. Indeed, we can rewrite our point estimate in  Equation~\ref{eq:quant-recall-using-estimated-reviewed-relevant} as an approximation of the expected value of a random variable:  
\begin{equation}
           \mathbb{E} \left[ 
                  \frac{
                    D_\mathcal{R}                  
                  }{ 
                    D_\mathcal{C}
                  }
                  \right]
           =
           \mathbb{E} \left[ 
                  \frac{
                   \sum_{i \in \mathcal{R}} D_i
                  }{ 
                    \sum_{i\in \mathcal{C}} D_i 
                  }
                  \right]
            \approx
            \frac{
                  \mathbb{E} \left[ \sum_{i \in \mathcal{R}} D_i \right]
                  }{ 
                   \mathbb{E} \left[ \sum_{i\in \mathcal{C}} D_i \right]
                  } 
              =
                   \frac{
                  \sum_{i \in \mathcal{R}} p_i
                  }{ 
                   \sum_{i\in \mathcal{C}} p_i
                  }
\end{equation}
where
$D_\mathcal{R}=\sum_{i\in \mathcal{R}} D_i$, $D_\mathcal{U}=\sum_{i\in \mathcal{U}} D_i$, and $D_\mathcal{C}=D_\mathcal{R} + D_\mathcal{U}$. 
(This expression also 
makes clear that our point estimate uses the ratio of estimated values of random variables to approximate the computationally awkward expected value of a ratio of random variables.)

Approximating $D_\mathcal{R} / (D_\mathcal{R}+D_\mathcal{U})$ by a Taylor series truncated to first order gives~\cite{papanicolaou2009taylor}:  
\begin{align}
& \phantom{{}+{}} Var\left( \frac{D_\mathcal{R}}{D_\mathcal{R}+D_\mathcal{U}} \right) \nonumber\\
    = & \phantom{{}+{}} 
        \E{ \frac{\partial f(D_\mathcal{R}, D_\mathcal{U})}{\partial D_\mathcal{R}} }^2 Var(D_\mathcal{R}) \nonumber 
      + \E{ \frac{\partial f(D_\mathcal{R}, D_\mathcal{U})}{\partial D_\mathcal{U}} }^2 Var(D_\mathcal{U}) \nonumber \\
    & + 2 \E{ \frac{\partial f(D_\mathcal{R}, D_\mathcal{U})}{\partial D_\mathcal{R}} }
        \E{ \frac{\partial f(D_\mathcal{R}, D_\mathcal{U})}{\partial D_\mathcal{U}} }  
        Cov(D_\mathcal{R},D_\mathcal{U}) 
\label{eq:recall-taylor-approx}
\end{align}
where
\begin{align}
\frac{\partial f(D_\mathcal{R}, D_\mathcal{U})}{\partial D_\mathcal{R}} 
    &= 
        \frac{1}{D_\mathcal{R}+D_\mathcal{U}} - \frac{D_\mathcal{R}}{(D_\mathcal{R}+D_\mathcal{U})^2} \label{eq:recall-partial-R} \\
\frac{\partial f(D_\mathcal{R}, D_\mathcal{U})}{\partial D_\mathcal{U}} 
    &= 
        \frac{-D_\mathcal{R}}{(D_\mathcal{R}+D_\mathcal{U})^2} 
        \label{eq:recall-partial-U}
\end{align}

Since the partial derivative of $D_\mathcal{R}$ (Equation~\ref{eq:recall-partial-R}) is always positive and the partial derivative of $D_\mathcal{U}$ (Equation~\ref{eq:recall-partial-U}) is always negative, the coefficient of the covariance $Cov(D_\mathcal{R},D_\mathcal{U})$ is negative. By omitting the negative terms, an upper bound on the variance is:  
\begin{align}
Var\left( \frac{D_\mathcal{R}}{D_\mathcal{R}+D_\mathcal{U}} \right) \le &\phantom{{}+{}}
      \E{\frac{1}{(D_\mathcal{R}+D_\mathcal{U})^2}} Var(D_\mathcal{R}) \nonumber \\
    &+ \E{\frac{D_\mathcal{R}^2}{(D_\mathcal{R}+D_\mathcal{U})^4}}
        \Big( Var(D_\mathcal{R}) + Var(D_\mathcal{U}) \Big) \nonumber
\end{align}
The logistic regression model is based on the assumption that the Bernoulli random variables are independent. Making that assumption here as well, and continuing to approximate the expected value of a ratio by a ratio of expected values, we can approximate the right hand side by

\begin{align}
    \frac{ Var(D_\mathcal{R}) }{(\widehat{R}_r+\widehat{U}_r)^2} 
    + 
    \frac{
        \widehat{R}_r^2 \left( Var(D_\mathcal{R}) + Var(D_\mathcal{U}) \right) 
    }{
        (\widehat{R}_r+\widehat{U}_r)^4
    }\nonumber
\end{align}
where 
\begin{align}
Var(D_\mathcal{S}) = Var\left( \sum_{i\in \mathcal{S}} D_i \right) = \sum_{i\in \mathcal{S}} Var( D_i )
       = \sum_{i\in \mathcal{S}} p_i(1-p_i)
\end{align}
for $\mathcal{S} = \mathcal{R}$ and $\mathcal{U}$. 

Finally, by assuming the recall estimator $D_\mathcal{R}/(D_\mathcal{R}+D_\mathcal{U})$ is approximately normally distributed, the 95\% confidence interval of the recall is 
\begin{align}
& \frac{\widehat{R}_r}{\widehat{R}_r + \widehat{U}_r} \\
\pm & 2 \sqrt{
\frac{1}{(\widehat{R}_r+\widehat{U}_r)^2} Var(D_\mathcal{R}) + \frac{\widehat{R}_r^2}{(\widehat{R}_r+\widehat{U}_r)^4}\Big( Var(D_\mathcal{R}) + Var(D_\mathcal{U}) \Big)
}
\end{align}

\subsection{Stopping Rules Based on Recall Estimates}

Given the above, we define two stopping rules based on the above approach of quantifying the number of relevant documents using model-based estimation. The \texttt{Quant} method stops the TAR process when the point estimate of recall (Equation~\ref{eq:quant-recall-using-estimated-reviewed-relevant}) first reaches or exceeds the recall goal.  The \texttt{QuantCI} method stops the TAR process when the lower end of a 95\% (2 standard deviations) confidence interval first reaches or exceeds the recall goal.  As the number of reviewed relevant documents increases, \texttt{QuantCI} approaches the behavior of \texttt{Quant}.

\section{Experimental Setup}

\subsection{Dataset}

We simulated recall-controlled retrieval tasks using RCV1-v2~\cite{rcv1}, a widely-used text categorization collection consisting of 804,414 news articles and 658 categories (including topics, industries, and geographical regions). 

To study tasks with a wide variety of characteristics, we selected the same 5 categories from each of 9 bins based on three ranges of class prevalence and three ranges of difficulties from the previous studies~\cite{cost-structure-paper}. 
Each selected category exhibits different characteristics, enabling the simulation of a wide variety of tasks. 
The lowest and highest prevalence bins were not used due to the wide range of values. Difficulty bins were based on the effectiveness of a logistic regression model trained on a random 25\% of documents and evaluated by R-precision on the remaining 75\%. 
For efficiency purposes, we downsampled the collection to 20\%. %

\subsection{Implementation}
\label{subsec:implementation}

We implemented active learning based on \texttt{libact}~\cite{libact}, an open-source, active-learning experimental framework. At round 0, one random positive document is used as the seed document to instantiate the active learning, and 200 documents based on relevance feedback~\cite{rocchio1971relevance, cormack2014evaluation} are sampled and reviewed at each round after.

Supervised learning used the \texttt{scikit-learn} implementation of Logistic Regression with an L2 regularization weight of 1.0. All words are used as features and are represented by BM25-saturated term frequencies~\cite{robertson2004understanding,yang2019sigir}. 

\subsection{Baselines}
We compare \texttt{Quant} and \texttt{QuantCI} with multiple existing heuristics stopping methods, which are all target-agnostic except the CMH heuristic method (\texttt{CMH-heuristic}). These target-agnostic rules include, the Knee and Budget Methods~\cite{cormack2016engineering}, \texttt{2399-rule}, \texttt{BatchPos}, \texttt{MaxProb}, and \texttt{CorrCoef}.

The Batch Positive Method (\texttt{BatchPos}) stops when the number of positive documents in the batch is less than or equal to 20 (i.e., precision is less than or equal 0.1) based on industrial convention. We tested stopping immediately (patient=1) and waiting for three additional rounds (patient=4) when the threshold is met.

The Probability Cutoff Method (\texttt{MaxProb}) stops when the maximum probability of the unreviewed documents is less than or equal to 0.1. This threshold yields a similar cost criterion as the \texttt{BatchPos} in which, in expectation, one positive document costs nine negative documents to retrieve. 

Apart from the rules discussed in Section~\ref{sec:existing-rules}, a score correlation-style approach that stops based on the correlation coefficients between two consecutive rounds (\texttt{CorrCoef}) is also tested. This approach stops when the average correlation coefficient of the last three rounds is greater than or equal to 0.99.

\subsection{Evaluation}
For each category, we run 10 active learning runs with different seed documents, with a total of 450 runs. The trajectories of the runs are fixed since the stopping methods tested do not intervene with the iterative process.

We evaluate the rules on the recall at stopping and the total cost at stopping for the given recall target. 

To quantitatively evaluate the reliability of each rule under different targets, we report the mean square error~(MSE) of the recall at stopping instead of the reliability measure (proportion of the runs that fulfilled the recall target at stopping) proposed by \citet{cormack2016engineering}.
The reliability measure favors stopping late since longer runs yield higher recall and, thus, easier to fulfill the target, resulting in massive recall overshooting for low recall targets. In an extreme case, a null stopping rule that never stops the process would have perfect but uninformative reliability for all recall targets. 
By reporting MSE, over and undershooting the target are both penalized.

We assign an \textit{idealized} cost penalty for undershoot the target recall~\cite{qbcb-paper}. If the process stops before fulfilling the target, we assign a cost penalty corresponding to the number of documents one needs to review based on the ordering suggested by the predictive model at stopping before reaching the target. This penalty is also called the optimal second phase review~\cite{cost-structure-paper}. If overshooting the recall, the cost of reviewing the excessive documents is also charged. For example, for a collection of 10,000 documents with 1,000 relevant documents and a recall target of 0.5, suppose one has only retrieved 300 relevant documents at iteration 10 (reviewed a total of 2,001 documents). If the rank list provided by the final predictive model requires going down to rank 2,300 among the 7,999 unreviewed documents to retrieve the additional 200 requested relevant documents, the idealized penalty is 2,300. And the total cost is therefore 4,301. 

In the next section, we normalize the cost of each run by its optimal cost. Since the cost varies by task characteristics and target, we report the ratio of the cost using a specific stopping rule over the optimal cost to avoid naturally high-cost categories diluting the values. In Table~\ref{tab:cost-ratio-table}, the average of such ratios over 450 runs is reported for each stopping rule. 
\section{Results and Analysis}

Both \texttt{Quant} and \texttt{QuantCI} demonstrate strong overall reliability on recall. In Table~\ref{tab:mse-recall}, \texttt{QuantCI} has the lowest MSE under the low recall targets (0.1 to 0.5) while the unadjusted version (\texttt{Quant}) remains strongly comparable. 
\texttt{QuantCI} is also very competitive in high recall targets (e.g., 0.7 and 0.9), where the Knee and Budget Methods are designed for. 
\texttt{CMH-heuristic} has an extremely small loss under 0.9 recall target but one of the worst in the lower recalls, indicating a vast overshooting.
In the next section, we focus on the distribution of recall at stopping points for each target-aware rule. 

The budget method provides the most stable recall over the 450 runs among the tested existing target-agnostic heuristic stopping rules in high recall targets. As shown in Table~\ref{tab:mse-recall}, it presents the smallest loss on 0.7 and 0.9 recall targets. The knee method has a slightly larger loss than the budget method. Both rules met their designed purpose for high-recall tasks but overshot for lower recall targets, such as 0.3, as they are unaware of the target. 

Other methods exhibit large variations across the tasks and runs, making these rules impractical for any recall target. The \texttt{BatchPos-20-1} stops extremely early, favoring low recall targets but still exhibiting a large loss. While \texttt{BatchPos-20-4} delays the stopping and reduces the recall loss under higher recall targets, the loss is still large compared to other approaches. 
Despite being target-agnostic, stopping rules that fail to provide a stable recall cannot serve as the stopping rule even for a specific recall target. 
In the rest of the section, we compare the target-aware methods with the Knee and Budget Methods. 

\begin{table}[]
\caption{Mean square error (MSE) of recall at stopping against different recall targets. The last column reports the mean of all targets of each rule. }
\label{tab:mse-recall}

\centering
\resizebox{\columnwidth}{!}{
\begin{tabular}{l|rrrrr|r}
\toprule
{} &  0.1 &  0.3 &  0.5 &  0.7 &  0.9 & Avg. \\
\midrule
Knee          &  0.522 &  0.279 &  0.115 &  0.032 &  0.029 &  0.195 \\
Budget        &  0.546 &  0.294 &  0.123 &  \textbf{0.031} &  0.019 &  0.203 \\
2399-Rule     &  0.493 &  0.272 &  0.131 &  0.070 &  0.089 &  0.211 \\
CorrCoef      &  0.445 &  0.241 &  0.117 &  0.072 &  0.108 &  0.197 \\
MaxProb-0.1   &  0.447 &  0.253 &  0.140 &  0.106 &  0.153 &  0.220 \\
BatchPos-20-1 &  0.294 &  0.199 &  0.185 &  0.250 &  0.396 &  0.265 \\
BatchPos-20-4 &  0.446 &  0.258 &  0.150 &  0.122 &  0.173 &  0.230 \\
\midrule
CMH-heuristics&  0.498 &  0.362 &  0.196 &  0.063 &  \textbf{0.008} &  0.225 \\
Quant         &  \textbf{0.236} &  0.164 &  0.102 &  0.072 &  0.050 &  0.124 \\
QuantCI       &  \textbf{0.236} &  \textbf{0.163} &  \textbf{0.092} &  0.046 &  0.021 &  \textbf{0.112} \\
\bottomrule
\end{tabular}
}
\end{table}

\subsection{Recall at Stopping}

\begin{figure*}
    \centering
    \includegraphics[width=\linewidth]{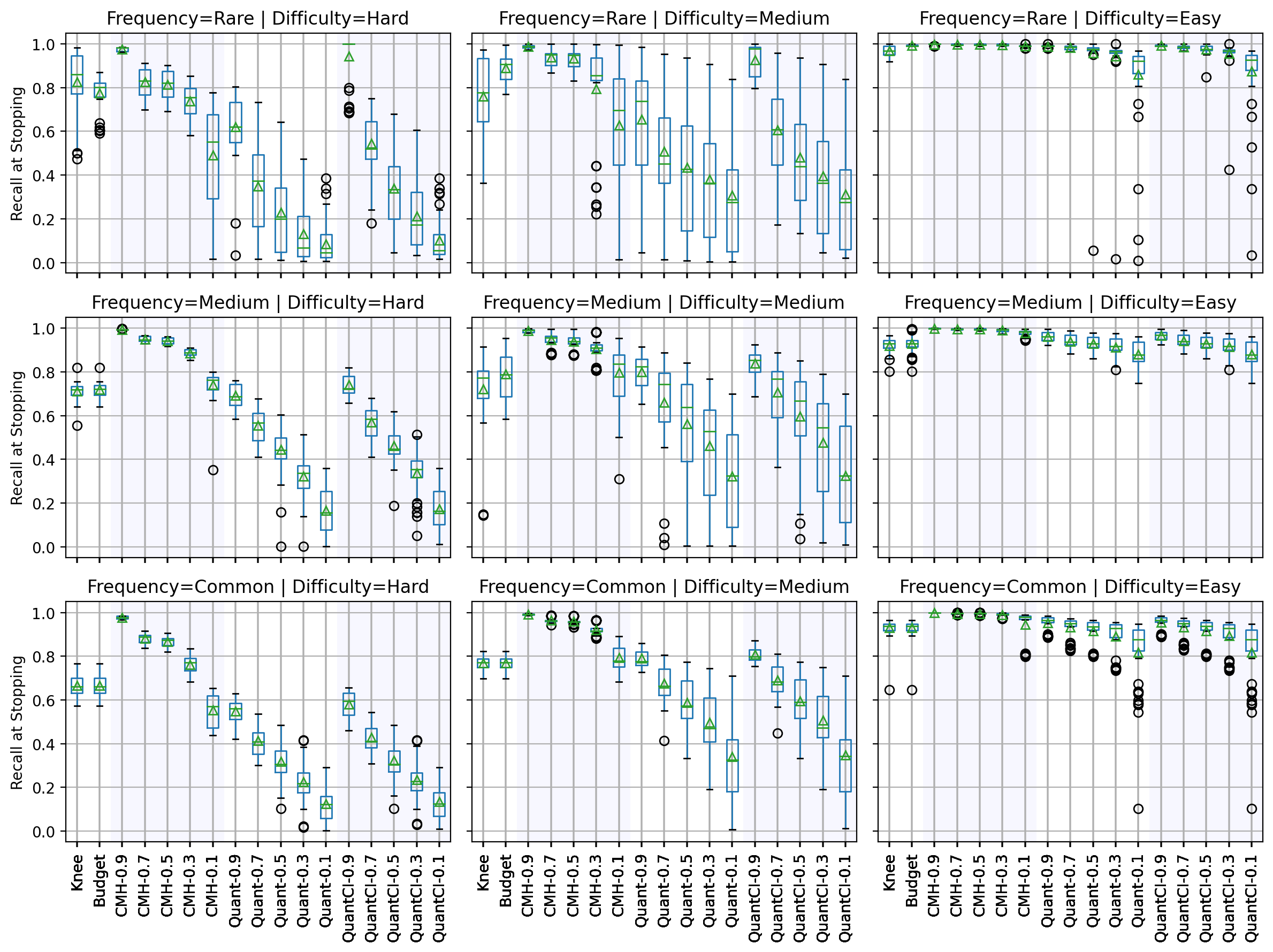}
    \caption{Recall at stopping of recall target-aware stopping rules. The numbers associated with the labels on the x-axis are the target recall. 50 runs (10 seeds and 5 categories) of each category bin are displayed using boxplot conventions: the box ranges from the 25\% (Q1) to 75\% (Q3) quartiles of recall with the 50 runs; the green lines and triangles are the median and mean recall, respectively, and whiskers extend to the lesser of the most extreme cost observed or to a distance 1.5(Q3 - Q1) from the edges of the box. Outliers are presented as dots above and below the whiskers.}
    \label{fig:recall-box-quants}
\end{figure*}

All four target-aware stopping rules adjust the stopping decision based on the recall target and exhibit clear decreasing trends in Figure~\ref{fig:recall-box-quants}. 
Despite the high classification effectiveness on the \textit{easy} tasks (the right column of Figure~\ref{fig:recall-box-quants}), it is difficult to estimate the number of relevant documents in general. All four approaches overshoot all levels of recall targets in the three \textit{easy} category bins.

On the other hand, \texttt{CMH-heuristic}, on average, overshoots the target in all nine bins, similar to what we observed in Table~\ref{tab:mse-recall}. Despite the awareness the target, assuming parts of the active learning queried documents are random creates the dependency to the task's characteristics. For \textit{hard} categories that active learning often experiences difficulties to find relevant documents from some iterations, those reviewing batches are similar to random, resulting in better-tracked stopping recall. The assumption rarely holds from the actual scenarios in other bins, producing larger overshooting in results. 

\texttt{Quant}
tracks the target recall closer than \texttt{CMH-heuristic}. 
Since \texttt{Quant} estimates the number of unreviewed relevant document by the sum of the predicted probabilities, the estimation is usually too big for \textit{easy} categories (a lot of documents seem positive) but too small for \textit{hard} categories (most documents seem negative), resulting in overshooting and undershooting the recall, respectively. The \textit{medium} categories (the middle column in Figure~\ref{fig:recall-box-quants}) strike a balance between the two and, therefore, allow the \texttt{Quant} to more accurately track the recall. 

Finally, 
\texttt{QuantCI} 
further improves the confidence of stopping and alleviates undershooting the target. 
\todo{OF:  Need to motivate QunatCI -- unless well motivated earlier}
For \textit{rare} categories (the first row in Figure~\ref{fig:recall-box-quants}), recall at stopping for the high recall targets increases while remaining similar recall for the lower recall targets. This phenomenon indicates that the standard deviation adjustment does not uniformly delay the stopping like the patient parameter for \texttt{BatchPos} but delays according to the quality of the estimation. 

Despite the similarity between \texttt{Quant} and \texttt{QuantCI} in the recall at stopping, the costs induced by each approach  significantly differ. In the next section, we discuss the total cost of applying each stopping rule. 

\vspace{-1em}
\subsection{Total Cost at Stopping}

\begin{table*}

\newcommand{\z}{\phantom{0}}
\newcommand{\cc}{\phantom{(00.00)}}
\newcommand{\cen}[1]{\multicolumn{1}{c}{#1}}

\newcommand{\sk}{${}^{\dagger}$}
\newcommand{\sq}{${}^{*}$}

\caption{Average ratio over the cost of stopping each category against the optimal cost for each recall target. Values in parentheses are the standard deviations. \sk~and \sq~indicate the statistical difference with 95\% confidence of \texttt{QuantCI} against the Knee Method and \texttt{Quant}, respectively. The significant test is conducted by paired t-test with Bonferroni correction. }
\label{tab:cost-ratio-table}

\centering
\begin{tabular}{l|rrrrr}
\toprule
{} &  \cen{Target = 0.1} &  \cen{Target = 0.3} &  \cen{Target = 0.5} &  \cen{Target = 0.7} &  \cen{Target = 0.9} \\
\midrule
Knee               &   21.52 (\z62.78) &   5.96 (10.07) &  3.43 (\z5.40) &  2.42 (\z3.13) &  \sk\textbf{2.36} (\z1.70) \\
Budget             &    56.45 (180.25) &  13.62 (43.61) &  5.43 (\z7.93) &  4.02 (\z6.70) &  3.35 (\z4.16) \\
2399-Rule          &   22.52 (\z41.64) &  7.16 (\z9.56) &  4.35 (\z3.99) &  3.76 (\z4.30) &  3.42 (\z4.36) \\
CorrCoef           &   16.64 (\z30.13) &  5.80 (\z7.06) &  3.94 (\z4.87) &  3.48 (\z6.06) &  3.64 (\z5.51) \\
MaxProb-0.1        &   11.29 (\z12.93) &  5.42 (\z4.12) &  4.93 (\z6.27) &  4.75 (\z7.61) &  3.56 (\z5.65) \\
BatchPos-20-1      &   12.44 (\z11.90) &  12.41 (13.94) &  13.90 (16.51) &  13.72 (18.48) &  10.34 (20.34) \\
BatchPos-20-4      &   12.89 (\z17.27) &  5.84 (\z4.39) &  5.36 (\z6.04) &  5.19 (\z7.40) &  3.57 (\z4.62) \\
\midrule
CMH-heuristics     &   27.37 (\z59.29) &  18.18 (30.47) &  16.18 (14.67) &  12.28 (13.21) &  17.33 (25.30) \\

Quant              &   6.12 (\z\z9.45) &  3.47 (\z3.60) &  3.07 (\z3.71) &  2.94 (\z3.53) & \sq2.81 (\z1.81) \\
QuantCI            &\sk\textbf{6.01} (\z\z9.79) &\sk\sq\textbf{3.01} (\z3.15) &\sk\sq\textbf{2.45} (\z2.41) &\sq\textbf{2.38} (\z1.95) &   3.35 (\z3.57) \\

\bottomrule
\end{tabular}
\end{table*}

Among the target-agnostic rules we tested, the Knee Method provides the lowest cost as shown in Table~\ref{tab:cost-ratio-table}. The Budget Method, despite stops with stabler recall, costs  significantly higher than the Knee Method under all tested recall targets (95\% confidence with Bonferroni correction of 5 tests on paired t-test). Even with 0.7 as the recall target, where the rules are designed for, the Budget Method (in contrast to the implication from the name) costs 4.02 times more than the optimal cost compared to 2.42 times by the Knee Method. 

For lower recall targets, such as 0.1, target-agnostic rules all induce large cost overheads. Since these heuristics rules all rely on certain notions of stability testing, the models are usually still unstable when the low recall targets are reached, resulting in vast overshooting and excessive reviewing cost. 

Simple rules such as \texttt{MaxProb}, \texttt{BatchPos}, and \texttt{2399-Rule}, while exposing higher cost in high recall targets, yield cost lower than the Knee Method. Since the stability tests of these simple approaches are weaker, they tend to stop the process earlier than the more sophisticated ones, such as Knee and Budget Methods. 

The awareness of the target does not guarantee lower cost. \texttt{CMH-heuristic} creates higher cost than \texttt{BatchPos-20-4} (95\% confidence with Bonferroni corrections of 5 tests on paired t-test). Regardless of the decreasing trend, \texttt{CMH-heuristics} still vastly overshoots the target. 
The average recall at stopping \texttt{CMH-heuristic} rule is 0.76 with a 0.1 recall target, which is even higher than 0.67 using \texttt{BatchPos-20-4} for all targets. 

On the other hand, both our proposed methods, \texttt{Quant} and \texttt{QuantCI}, tightly track the target to provide a total cost that is closer to optimal. 

For low (e.g., 0.1) to medium (e.g., 0.5) recall targets, \texttt{QuantCI} yields significantly lower cost than any other tested rule except \texttt{Quant} with 0.1 recall target (95\% confidence with Bonferroni corrections of 27 tests on paired t-test). 
In Figure~\ref{fig:recall-box-quants}, the trend of recalls at stopping using \texttt{Quant} is slightly concave or convex in some category bins, such as \textit{Rare-Hard} and \textit{Medium-Medium}. Since the \texttt{Quant} method estimates the total number of relevant documents based on the probabilities produced by the classifiers, the quality of such estimations depends on the task's difficulty. Therefore, standard deviation adjustment provides additional protection to the stopping rule against the instability that often appears during early iterations of active learning. The resulting recall at stopping approaches linear in Figure~\ref{fig:recall-box-quants}. The cost also reduces significantly.

However, for 0.1 recall target, the optimal stopping point (iteration just achieved the target) often appears when the predicted probabilities are not yet stabilized. For \texttt{QauntCI}, it conservatively waits until the predictions are sufficiently confident  (when the lower bound of confidence intervals are larger than the target), usually implying overshooting the target. Therefore, despite that the average cost ratio of \texttt{QuantCI} is lower than \texttt{Quant} under 0.1 recall target, the variance of \texttt{QuantCI} is larger, and the difference is not significant. 

For higher recall targets (e.g., 0.7 and 0.9), our proposed approaches perform similarly with the Knee Method, the best existing heuristic rule for such targets. Note that the Knee Method was designed for a 0.7 recall target. Our proposed \texttt{Quant} and \texttt{QuantCI} produce very similar cost even in this crowded battlefield. 

When targeting 0.9 recall, the Knee Method offers a slightly lower cost ratio than \texttt{Quant} and \texttt{QuantCI}. Again, since the \texttt{Quant} method relies on the underlying classifier to estimate the total number of relevant documents, identifying the last several relevant documents is challenging both ranking-wise (for prioritizing the reviewing) and probabilistic-wise (for estimating its existence). 
When the last several needles are still in the haystack, the number of relevant documents in the unreviewed set is under-estimated, resulting inflated estimated recall, and therefore, undershooting the real recall. 
Furthermore, the variance of such recall estimation during those unstable stages is undoubtedly high, driving \texttt{QuantCI} to stop the process later; often, it is too late and overshoots the recall. This results in cost overheads (3.35). 

Both \texttt{Quant} and \texttt{QuantCI} outperform the widely-compared target-agnostic Knee Method except when setting an extremely high recall target without any intervention and additional random samples. 
The challenging high recall target often requires a high reviewing cost even with optimal stopping, providing a large incentive for using a sample-based stopping approach. The cost of the random sample can be amortized to the longer active learning run. Next, we investigate a specific active learning run closely.

\subsection{A case study on stopping and cost}

\begin{figure*}[t]
    \centering
    \includegraphics[width=\linewidth]{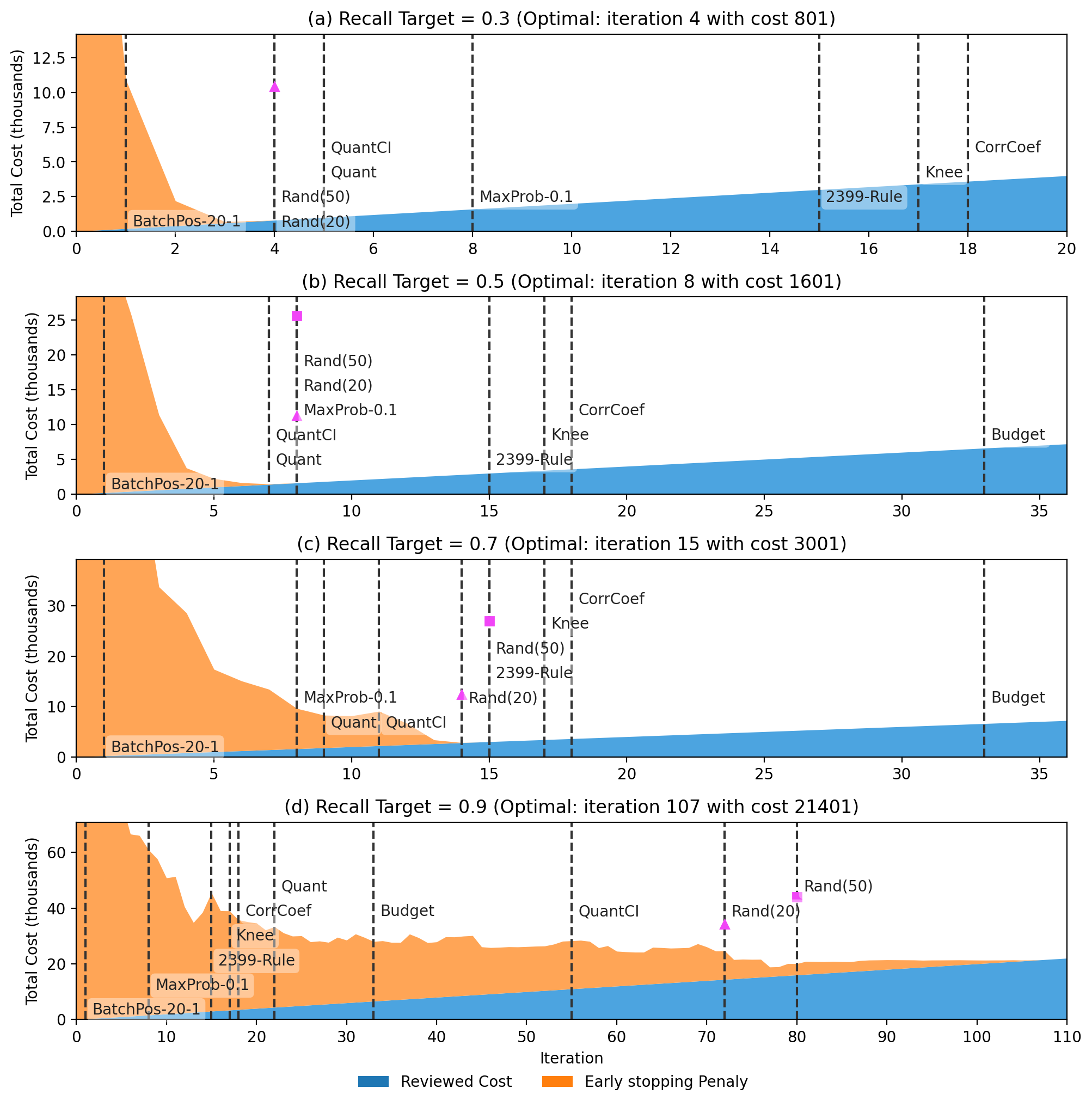}
    \vspace{-1.5em}
    \caption{Cost and stopping round of each rule on category \texttt{I24000} with seed \#7. The height of the blue and orange blocks indicate the cost and penalty if stopping at the corresponding round. Vertical dashed lines indicate the stopping round of the rules labeled at the right of the line. \texttt{CMH-heuristic} stops beyond the range presented for all recall targets. Triangles and squares represent the cost at stopping of random sample with 20 and 50 positive documents, respectively. }
    \label{fig:cost-dynamic-stopping}
\end{figure*}

In Figure~\ref{fig:cost-dynamic-stopping}, we demonstrate an active learning run on category \texttt{I24000} (a \textit{Rare}-\textit{Medium} category with prevalence 0.002133) with seed \#7. Each target-agnostic rule stops at the same iteration across the targets but appears at different locations because of each graph's different x-axis range. The optimal stopping point is where the early stopping penalty first becomes zero. 

Additional to the heuristic rules, we tested a sample-based rule that stops when the recall estimated by the sample reaches the target, which is similar to the Target Rule proposed by \citet{cormack2016engineering}.  We evaluated both samples with 20 and 50 relevant documents. By sampling uniformly at random until enough relevant documents are selected, the expected sample sizes (including the non-relevant ones) are 9,378 and 23,445, respectively. We conducted 100 trials for each sampling size and reported the median stopping iterations. 

Despite the same increasing rate of the reviewing cost (blue blocks in Figure~\ref{fig:cost-dynamic-stopping}), stopping late for low recall targets with low optimal cost yields a higher relative cost overhead. This observation also implies that it is \textit{cheaper} to stop later for a high recall target than a low one. The same observation can be made on the left side of the optimal stopping point but with a convex relationship instead of linear due to the nonlinearity of the learning curves. If stopping vastly prematurely, e.g., \texttt{BatchPos-20-1}, the cost overhead is excessive because of the under-trained model. For the 0.9 recall target, the improvement of the model plateaued (the decrements of the orange block), so stopping between iteration 25 to 70 yields a very similar total cost.  

Besides leading to a lower cost, the overhead of stopping early and late differs across recall targets as well. The shape of the penalty also differs based on the characteristics of the task. Due to the space constraint, we cannot demonstrate other categories. But the distinctions between the targets already demonstrate the need to track the target. 

Compared to the heuristic approaches, estimating the progress based on a random sample is much more accurate than heuristics. In each tested recall target except 0.9, the sampling approach either stops at the optima or off by one iteration. However, this accuracy comes with a price on the sample, which in total cost usually exceeds heuristic approaches.
For the challenging 0.9 recall target, the sampling approaches are still much more accurate than the heuristics but are off by more iterations. In this case, the cost of using such sampling methods is similar or lower than the heuristic ones, providing the incentive of reviewing the large random sample a priori. 

However, routinely testing the same goal under the same random sample yields sequential biases~\cite{webber2013approximate}. The simple approach we tested here is no exception. While not proposing this simple rule for a high recall target, this observation motivates the usage of sample-based stopping rules. A valid approach that avoids sequential biases is still an active research topic.

\section{Summary}

In this work, we proposed a target-aware heuristic stopping rule \texttt{Quant} that estimates the total number of relevant documents and the recall progress using a model-based approach for one-phase technology-assisted review. We provided the standard deviation of our recall estimator and further improved the stopping rule based on the stability of the estimation. The \texttt{Quant} Method with standard deviation adjustment \texttt{QunatCI} has the lowest MSE on recall at stopping among the tested rules. It also yields the lowest total cost when targeting low recall and competitive with the popular Knee method for high recall targets. 

\bibliographystyle{ACM-Reference-Format}
\bibliography{ms}

\end{document}